# A physiological model of the inflammatory-thermal- pain-cardiovascular interactions during a pathogen challenge

**Atanaska Dobreva[1,*], Renee Brady[2,*], Kamila Larripa[3], Charles Puelz[4], Jesper Mehlsen[5], Mette S. Olufsen[1,#]**

[1]Department of Mathematics, North Carolina State University, Raleigh, NC

[2]Department of Integrated Mathematical Oncology, H. Lee Moffitt Cancer Center and Research Institute, Tampa, FL

[3]Department of Mathematics, Humboldt State University, Arcata, CA

[4]Courant Institute of Mathematical Sciences, New York University, New York, NY

[5] Section for surgical pathophysiology, Rigshospitalet, Copenhagen, Denmark

**Running title: Inflammatory-thermal-pain-cardiovascular response to a pathogen challenge**

[*]Atanaska Dobreva and Renee Brady are joint first authors.

[#]**Corresponding author**

Mette Olufsen

SAS Hall 3216

Department of Mathematics, Box 8205

North Carolina State University

Raleigh, NC 27695

Email: msolufse@ncsu.edu

Phone: 919-515-2678

## Abstract

Uncontrolled, excessive production of pro-inflammatory mediators from immune cells and traumatized tissues can cause systemic inflammatory issues like sepsis, one of the ten leading causes of death in the United States and one of the three leading causes of death in the intensive care unit. Understanding the effects of inflammation on the autonomic control system can improve a patient's chance of recovery after an inflammatory event such as surgery. Though the effects of the autonomic response on the inflammatory system are well defined, there remains a gap in understanding the reverse response. Specifically, the impact of the inflammatory response on the autonomic control system remains unknown. In this study, we investigate hypothesized interactions of the inflammatory system with the thermal and cardiovascular regulatory systems in response to an endotoxin challenge using mathematical modeling. We calibrate the model to data from two independent studies: a) of the inflammatory response in healthy young men and b) a comparative study of the inflammatory response between mice and humans. Simulation analysis is used to explore how the model responds to pathological input and treatment, specifically antibiotics, antipyretics, vasopressors, and combination therapy. Our findings show that multimodal treatment that simultaneously targets both the pathogen and the infection symptoms gives the most favorable recovery outcome.

## New & Noteworthy

This study introduces an innovative computational model incorporating the behavior and interactions of inflammation, temperature, pain, HR, and BP in response to an endotoxin challenge. We use sensitivity analysis and subset selection to identify parameters that can be estimated given the model and data and use nonlinear least squares optimization to calibrate the model to data from two independent clinical studies. Optimal parameters are used to predict how patients may respond to antibiotics, antipyretics, vasopressors, and a combination therapy. Results of our study reveal that the inflammatory response to a pathogen challenge elevates temperature, causing an increase in HR on a timescale of hours, and that temperature plays a vital role in the interaction between BP and HR. Moreover, we show that during an infection, treatment with antibiotics, antipyretics, vasopressors, and combinations of these acts to down-regulate HR in a temperature-dependent fashion.

**Abbreviations**

(In order of appearance)

| | |
|---|---|
| LPS | Lipopolysaccharide |
| CNS | Central nervous system |
| NO | Nitric oxide |
| HRV | Heart rate variability |
| PT | Pain threshold |
| BP | Blood pressure |
| HR | Heart rate |

**Introduction**

Systemic inflammation is comorbid with various diseases including diabetes (64), cancer (14), heart disease (66), and sepsis (29, 30), and it plays a prominent role in eight out of the ten leading causes of death in the U.S. (28). The action of inflammation is multi-faceted and impacts multiple organ systems, such as the digestive, respiratory, endocrine, and nervous systems. In this study, we use mathematical modeling to explore how the inflammatory response to a pathogen challenge interacts with the thermal, pain perception, and cardiovascular systems. Our goal is to examine the effect on cytokine concentrations, body temperature, pain, heart rate (HR), and blood pressure (BP). We focus on hemodynamic quantities that are routinely collected at both in- and out-patient facilities, measured non-invasively at a high sampling rate, and interpreted within minutes. We hypothesize that the hemodynamic signals correlate with inflammatory markers, and therefore have the potential to inform the development of clinical protocols important for early detection of pathological conditions such as sepsis. Understanding how hemodynamics change with inflammation is also of great importance in devising more effective treatment strategies to improve recovery outcomes for patients.

Numerous *in-vivo* biological studies in mice and rats have investigated the effects of autonomic control on the inflammatory system (56, 57). Borovikova *et al.* (5) discovered the cholinergic anti-inflammatory pathway modulating the immune response via the local release of acetylcholine from vagal fibers in target tissues. They found that an increase in vagal activity results in a decrease in the pro-inflammatory cytokines tumor necrosis factor-$\alpha$ (TNF-$\alpha$), interleukin 6 (IL-6), and interleukin 1$\beta$ (IL-1$\beta$). Following this, Tracey (57) found that the production of cytokines in response to inflammation activates afferent firing to the brain, and that subsequent vagal efferent activation inhibits cytokine synthesis. This inflammation-sensing and inflammation-suppressing network, dubbed the inflammatory reflex, explains how the autonomic control and inflammatory systems work synergistically to maintain homeostasis of the inflammatory response. The study by Tracey (57) did an excellent job describing the impact of the autonomic control on inflammation, but it did not describe how changes in autonomic regulation, triggered by inflammation, impact hemodynamics.

Previous studies have used modeling to explore the inflammatory response to an endotoxin challenge in rodents in conjunction with *in-vivo* experimental data. Seminal studies include contributions by Chow *et al.* (10) who modeled immunocytes, nitric oxide (NO) byproducts, and three distinct cytokines to investigate the acute inflammation response to a bolus administration of the bacterial endotoxin lipopolysaccharide (LPS). The Chow *et al*. study demonstrated that three different triggers (infection, surgery, and hemorrhage) engage a universal cascade of immune signals, whose behavior and magnitude of activity differ among the shock states. A similar model by Daun *et al*. (15) studied how different tissue damage levels in rats affect cytokine behavior, and Prince *et al*. (47) examined the trauma-induced inflammatory pathway of mice deficient in CD14, a molecule in immune cell receptors used for LPS recognition.

While the above investigations (10, 15, 47) revealed essential features of the immune reaction to pathogens in rodents, they do not elucidate how the inflammatory system interacts with the cardiovascular, thermal, and pain systems. Moreover, it is well known that the inflammatory pathways differ between rodents and humans (54) and that rodents tolerate much higher levels of pathogens (17). Experimental observations by Copeland *et al*. (13) demonstrated this by administering LPS to humans and mice at doses (2 ng/kg of body weight for humans and 500 ng/kg for mice) eliciting equivalent IL-6 plasma concentrations 2 hours post-injection. Their results showed that humans experienced fever as well as an increase in HR and systolic BP, while the mice had no fever or changes in HR and BP. These results demonstrate that investigations in mice do not have direct applicability to humans. New models are needed to effectively explore the differences between the two species and to gain insight into thermal regulation and cardiovascular functions.

Several studies have used models to examine inflammatory-cardiovascular communication in humans (18, 52). Foteinou *et al*. (18) employed a multi-scale mathematical model to predict the parasympathetic activity and HR in human subjects who received a dose of LPS (2 ng/kg of body weight) alone or in combination with an epinephrine infusion. This study assessed parasympathetic activity via calculations of the heart rate variability (HRV) index $pNN_{50}$, which measures the percentage of differences of successive inter-beat intervals greater than 50 ms (18). Scheff *et al*. (52) expanded this model by incorporating regulation of immune activity via hormonal circadian rhythms. They showed that circadian variability in inflammation correlates with daily patterns in HR and HRV (52). While the investigations of Foteinou *et al*. (18) and Scheff *et al*. (52) reflect the complexity of inflammatory-cardiac interactions, they do not account for thermal and pain regulation and the impact on hemodynamics.

As the body battles infection the threshold for pain perception decreases (2, 13, 31, 63), and the release of inflammatory cytokines, in particular, IL-6 and TNF-$\alpha$, raises core temperature (61). Results by Copeland *et al*. (13) showed that during a pathogen challenge in humans, the change in BP and temperature plays a vital role in HR modulation. Septic patients typically develop a high fever, low BP and high HR (40). Considering this evidence, we hypothesize that changes in temperature and pain perception observed in humans contribute to the fluctuations in HR and BP following an endotoxin challenge.

The objective of this study is to develop a mathematical model to understand the impact of variations in human inflammatory markers on thermal, pain, and hemodynamic responses. We hypothesize that LPS-induced inflammation causes fever, leading to higher HR via decreased parasympathetic activity and changes in BP-HR interactions. In addition, we hypothesize that inflammation lowers the threshold for pain perception and stimulates NO synthesis. Finally, we hypothesize that pain and NO have opposing effects on BP. More specifically, a reduced pain tolerance increases BP through increased vascular resistance, and an increase in NO leads to a decrease in BP through vasodilation.

To test these interactions, we expand upon our previous dynamic model for a systemic inflammatory reaction to LPS (7) by adding pathways for changes in temperature, pain perception, NO, HR, and BP. The model is fit to patient data from two independent clinical studies (13, 31). We further perform simulation studies of the model's ability to capture the effects of common treatments, such as antibiotics, antipyretics, and vasopressors.

## Materials and Methods

## Data and Experimental Design

*Study Subjects*

Copeland *et al*. (13) administered a bolus dose of LPS to 10 human subjects. The participant group included male and female healthy adult volunteers aged 18 to 40 years. Exclusion criteria for participation included chronic disease history, e.g. cancer, rheumatoid arthritis, heart disease, and hypertension, or history of abnormalities affecting the immune system, e.g. the liver, kidneys, endocrine system, or neural system. This study was approved by the Institutional Review Board of UMDNJ-Robert Wood Johnson Medical School, all subjects gave written consent to participate. The study by Janum *et al*. (31) included 20 healthy male adults ages 18 to 35 years, who received LPS. Exclusion criteria include smoking, obesity, daily intake of medication, and splenectomy. The Regional Committee on Health Research Ethics and the Regional Data Monitoring Board approved the investigation protocol, and all participants gave written consent to participate in the study.

*Experimental Protocol*

Both studies followed a similar experimental protocol administering a 2 ng/kg dose of LPS derived from *Escherichia coli* intravenously and measuring the systemic inflammatory response, body temperature, HR, and BP. Copeland *et al*. (13) assessed participants over 9 hours post-injection, while Janum *et al*. (31) assessed subjects 2 hours before and 6 hours after endotoxin administration. The Janum study applied a heat stimulus to the non-dominant thigh at varying temperatures (45, 46, 47, 48° C) for 5 seconds followed by a period of 32° C to evaluate pain perception. Subjects were asked to rate each heat stimulus on a scale from 0 (*no pain*) to 10 (*worst imaginable pain*). Pain perception was measured using an algometer in which pressure was applied manually at increasing kPa. When participants indicated that their pain threshold was reached, the algometer value was recorded.

In each study, blood samples were collected to measure plasma levels of pro- (TNF-$\alpha$, IL-6, and IL-8) and anti-inflammatory (IL-10) cytokines. Both studies recorded systolic BP and temperature at discrete time points; standard BP was measured hourly with a BP cuff on the upper arm. The Copeland study measured HR at discrete time points, while Janum *et al*. (31) measured HR continuously throughout the 6-hour interval. Fig. 1 shows the detailed experimental protocol for each study. In our previous study (7) modeling the inflammatory response using the data by Janum *et al*. (31), results are reported as the mean $\pm$ standard error of the mean (SEM) for each subject. The data extracted from the Copeland *et al.* study were reported as the average over the population. To compare the response from the two studies we validated model predictions against the population average response for each study. Moreover, to predict proper cytokine decay synthetic data was added to the Janum *et al*. (31) data at $t = 7$ and 8 hours. Finally, to achieve data format consistency for BP and HR between the two studies, we sampled HR and BP measurements at the same time-points.

**FIG. 1 – DATA DESCRIPTION**

**Mathematical Model Development**

Our mathematical model (shown in Fig. 2) includes three sub-models of 1) the inflammatory response to endotoxin, 2) the effect on temperature, pain perception, and nitric oxide, and 3) their effect on the cardiovascular system. Below we describe each component and their interdependencies.

## FIG. 2 – FEEDBACK DIAGRAM FOR HUMAN RESPONSE TO LPS

*Part I: The inflammatory response to an endotoxin challenge*

Monocytes and macrophages use cytokine signaling to communicate in response to a pathogen and are an essential part of innate immunity. Bone marrow stem cells differentiate into monocytes and move into the bloodstream. From here they enter the connective tissue matrix where they differentiate into macrophages, which interact with the cytokines (44). Monocytes and macrophages are normally at rest, but with a bacterial stimulus (e.g. from LPS) the number of macrophages increases by several orders of magnitude.

Cytokines are potent signaling molecules that regulate many processes essential to immunity and inflammation. In (7), we constructed a classic kinetic model of the systemic inflammatory response to an endotoxin challenge, incorporating signaling pathways illustrated in Fig. 3 (equations are given in the Appendix). We extend this model to include thermal, pain, nitric oxide, and hemodynamic models and calibrate it to inflammatory mediator data from the studies by Janum *et al*. (31) and Copeland *et al*. (13).

## FIG. 3 – INTERACTIONS AMONG IMMUNE COMPONENTS

*Part II: The thermal, pain and nitric oxide response*

<u>*Thermal effects.*</u> The binding of LPS to receptors on macrophages and other immune cells stimulates the production of pyrogenic (fever-inducing) cytokines IL-6, TNF-$\alpha$, and IL-1$\beta$, which act to induce and maintain fever. These pyrogens stimulate afferent vagal nerves terminating in the brain, and this relays information to the hypothalamus about the ongoing inflammation and triggers the release of prostaglandins. Prostaglandins act on neurons in the preoptic nucleus of the hypothalamus, which is the area responsible for temperature regulation (12, 36). TNF-$\alpha$ is a potent pyrogenic cytokine, and it is one of the main cytokines implicated in septic shock. IL-6 has recently been shown to be necessary to sustain fever (16, 24). Conversely, IL-10 has antipyretic effects, lowering temperature (12, 43).

To account for this feedback, we introduce temperature regulation modeled as a function of TNF-$\alpha$ and IL-6, which upregulate temperature, and IL-10, which counteracts these effects. Temperature is modeled as

$$\frac{d\text{Temp}}{dt} = \frac{1}{\tau_1}(-\text{Temp} + T_b + k_T(T_M - T_b)(k_{T\text{TNF}}H_T^U(\text{TNF} - w_{\text{TNF}}) + k_{T6}H_T^U(\text{IL6} - w_{\text{IL6}}) - k_{T10}(1 - H_T^D(\text{IL10} - w_{\text{IL10}})))), \tag{3}$$

where Temp is temperature, $T_b$ and $T_M$ are the respective baseline and maximum temperatures, and $\tau_1, k_T, k_{T\text{TNF}}, k_{T6}, k_{T10}$ are rate constants. The baseline values of the cytokines are given by $w_X$, for $X \in \{\text{TNF-}\alpha, \text{IL6}, \text{IL10}\}$. The up- and down-regulation of Y by X is described by the equations $H_Y^U(X) = \frac{X^h}{\eta_{YX}^h + X^h}$ and $H_Y^D(X) = \frac{\eta_{YX}^h}{\eta_{YX}^h + X^h}$, respectively. The half-saturation is given by $\eta_{YX}$ and the exponent $h$ regulates the steepness of the Hill function.

While we recognize the importance of IL-$1\beta$ in the thermal response, we refrain from including it in the temperature equation because this cytokine was not measured in the clinical studies from which data was obtained.

*Pain Threshold (PT).* Sensory neurons at the site of infection are stimulated in the presence of inflammation and alter signaling to the CNS through vagal afferent fibers (35). Nociceptors are essential mediators of pain perception (45) and have nerve endings terminating near macrophages and other immune cells. Nociceptors express receptors for cytokines such as TNF-$\alpha$ and IL-6 (8), but they can also be activated by bacteria directly. Previous studies (2, 31, 63) show that there is a dose-dependent relationship between pain perception and inflammation. We model the threshold for perception of pain as

$$\frac{d\text{PT}}{dt} = -k_{\text{PTE}}E\,\text{PT} + k_{\text{PT}}(PT_b - \text{PT}), \tag{4}$$

where $k_{PTE}$ and $k_{PT}$ are rate constants and $PT_b$ is the baseline pain threshold. Upon injection, the endotoxin (E) decreases exponentially, resulting in a decrease in the pain perception threshold (59). As the body clears endotoxin, the pain perception threshold returns to baseline.

*Nitric Oxide (NO).* The endothelium cells lining the blood vessels uses nitric oxide (NO) signaling to interact with nearby smooth muscle cells, which relax with increased release of NO (37). NO has direct and indirect microbial effects including inhibition of pathogen proliferation (3, 4, 58). As part of the inflammation pathway, upon signaling from TLRs or inflammatory cytokines, macrophages produce inducible nitric oxide synthase (iNOS) which affects their phenotype and leads to the production of NO (39). TNF-$\alpha$ upregulates the NO production (51) and IL-10 downregulates its synthesis (9, 25). We model cytokine-mediated NO dynamics as

$$\frac{dN}{dt} = k_{NM} M_A \left( \frac{\text{TNF}(t-\kappa)^{h_{NTNF}}}{\text{TNF}(t-\kappa)^{h_{NTNF}} + \eta_{NTNF}^{h_{NTNF}}} \right) \left( \frac{\eta_{N10}^{h_{N10}}}{\text{IL}10(t-\kappa)^{h_{N10}} + \eta_{N10}^{h_{N10}}} \right) - k_N N. \quad (5)$$

In this equation $\kappa$ represents the delay in activation/inhibition of NO from TNF-$\alpha$ and IL-10, $k_{NM}$ and $k_N$ are rate constants, and $\eta_{NTNF}, \eta_{N10}, h_{NTNF},$ and $h_{N10}$ are Hill function parameters determining the half-saturation value and steepness of the response.

*Part III: Impact of the immune response on cardiovascular dynamics*

The cardiovascular system is continually modified to maintain homeostasis (ensuring adequate oxygen perfusion at stable resting BP). The body maintains this state via the autonomic control system regulating vascular compliance, resistance, and cardiac contractility and HR. The baroreflex branch of the autonomic control system consists of parasympathetic and sympathetic signaling primarily responding to changes in BP. The vagal nerve is the primary pathway for the parasympathetic signaling. An increase in parasympathetic signaling leads to a decrease in HR (62). This response is caused by release of acetylcholine primarily at the level of the sinoatrial node. Preganglionic sympathetic nerve fibers travel through the spinal cord synapses with the postganglionic fibers and thus transmit the sympathetic signals, stimulating the release of noradrenaline, which causes smooth-muscle contraction, increasing the peripheral vascular resistance and decreasing the compliance of the vascular wall. The baroreflex response to changes in BP acts within seconds to minutes, but basal activity in both efferent pathways (parasympathetic and sympathetic) are present even in the absence of a stimulus (change in BP). We hypothesize that the basal activity level may be modulated by several factors including fever and pain perception (21, 32). In addition to neural response, the vascular system is also controlled locally, e.g., in response to changes in NO, which is a potent vasodilator and through release of catecholamines form the adrenal medulla. To model this feedback, we develop a simple cardiovascular model coupled with control for modulating HR and peripheral vascular resistance, impacting prediction of BP.

*Cardiovascular Model*

Due to significant BP and HR changes taking place over the course of hours, the current model does not account for respiratory dynamics. Therefore, we limit the cardiovascular model to the systemic circulation, which we model as a series of compliance (C) and resistance (R) elements using an electric circuit analogy. Our model (shown in Fig. 4) consists of four compliance compartments representing the large and small arteries and veins. Three resistance elements separate these compartments, including the resistance provided by the systemic organs and the pressure gradient between the large and small arteries and veins. The model is non-pulsatile since the relevant timescale for all model variables is hours, i.e. the pumping of the heart is not explicitly incorporated (65).

The predicted variables are flow ($q \ mL/s$), volume ($V \ mL$), and pressure ($p \ mmHg$). The model is formulated using four differential equations describing conservation of volume via

$$\frac{dV_i}{dt} = q_{in} - q_{out},\qquad(6)$$

where $q_{in}$ is flow entering the compartment, and $q_{out}$ is flow leaving the compartment. Between two compartments, flow is related to pressure via Ohm's law given by

$$q_i = \frac{p_{out} - p_{in}}{R_i},\qquad(7)$$

where $p_{in}$ and $p_{out}$ (mmHg) denote the pressure in the two surrounding compartments and $R_i$ (mmHg s/mL) is the resistance to flow.

Finally, for each compartment, pressure ($p$) is related to volume ($V$) via a pressure/volume relation

$$p - p_{tis} = E(V - V_{un}),\qquad(8)$$

where $p_{tis}$ (mmHg) is the tissue pressure, $E$ is the elastance, and $V_{un}$ is the unstressed volume. We drive the cardiovascular model by a "non-pulsatile heart" tracking stroke volume $V_{str}$ via

$$Q \approx HV_{str},\qquad(9)$$

where $H$ (beats/min) is the HR, and $Q$ (mL/s) is the cardiac output. $V_{str}$, the volume of blood the heart pumps out during one beat, is computed by

$$V_{str} = V_{ED} - V_{ES} = -\left(\frac{p_{la}}{E_M} - \frac{p_{lv}}{E_m}\right),\qquad(10)$$

where $V_{ED}$ is the end-diastolic volume in the heart, $V_{ES}$ is the end-systolic volume, $p_{la}$ is pressure in the large arteries, $p_{lv}$ is pressure is pressure in the large veins, and $E_M$ and $E_m$ are the maximum and minimum elastance, respectively (1). The Appendix gives a complete list of model equations.

## FIG. 4 – CARDIOVASCULAR MODEL

_Cardiovascular Control Model._ As noted earlier, the cardiovascular control system modulates several vascular parameters including peripheral vascular resistance ($R_S$), cardiac contractility (here represented by $E_m$ and $E_M$, vascular compliance ($C$), and heart rate ($H$). Given the lack of pulsatilility in the model and data, we focus on describing regulation of vascular resistance and heart rate. In this system, vascular resistance ($R_S$), on the timescale studied here, is primarily controlled by pain via sympathetic stimulation, which up-regulates $R_S$, and nitric oxide which down-regulates $R_S$. In response to LPS, the pain perception threshold decreases, upregulating the peripheral vascular resistance between the small arteries and veins via sympathetic stimulation. Resistance is downregulated by NO. To capture these effects, we model the peripheral vascular resistance as

$$\frac{dR_s}{dt} = k_{RPT} \frac{\Gamma^2}{\Gamma^2 + \eta^2{}_{RPT}} - k_{RN}N - k_R(R_s - R_b), \qquad (11)$$

where $\Gamma = \frac{dPT}{dt}$ is the rate of change of the pain perception threshold and $R_b$ is the baseline peripheral vascular resistance (before the LPS injection) in the absence of pain and nitric oxide ($\Gamma = 0$ and $N = 0$). The rate constants are given by $k_{RPT}, k_{RN}$, and $k_{RN}$. The half-saturation value of the Hill function is given by $\eta_{RPT}$ Upon administration of endotoxin, the pain threshold decreases, resulting in an increase in $\Gamma^2$ and consequently $R_s$. As LPS decays and PT returns to its baseline value, $\Gamma^2$ approaches zero. NO begins to rise two to four hours after the initial inflammatory response, causing $R_s$ to decrease.

The changes in peripheral vascular resistance lead to changes in BP. Elevated vascular resistance, brought about by stimulation of the sympathetic system in response to increased pain perception, leads to a rise in BP. On the other hand, a high NO concentration notably lowers resistance through vasodilation, which can lead to an excessive drop in BP (50).

Due to activation of the baroreflex, the increase in BP leads to a decrease in HR, while during hypotension, the excessive BP drop causes an increase in HR. Also, with fever induction as a consequence of the pyrogens released by monocytes, the CNS decreases basal vagal tone, which increases HR (32). To predict these responses in HR dynamics, we model HR as follows

$$\frac{dH}{dt} = \frac{-H + k_H(H_M - h_b)H_H^U(\text{Temp} - T_b)f(BP, BP_b) + H_b}{\tau_2}, \tag{18}$$

where

$$f(BP, BP_b) = \begin{cases} H_H^U(BP_b - BP), & \text{if } BP \leq 100 \text{ mmHg} \\ H_H^D(BP - BP_b), & \text{if } BP > 100 \text{ mmHg} \end{cases}.$$

We prescribe a switching systolic BP level of 100 mmHg since the hypotension cutoff value can range between 90 mmHg and 117 mmHg, depending on age (42). HR will increase in response to hypotension, which is supported by the observation that when BP is low HR rises to compensate for reduced cardiac preload resulting from vascular dilation (23). Otherwise, BP would act to lower HR. Both effects depend on temperature elevated above its baseline level. In sum, we are proposing that BP impacts HR on the time scale of hours during infections, but not in the absence of fever under normal circumstances.

**Model Summary**

The model described above can be written in the following form

$$\frac{dx}{dt} = f(x, t; \theta), \tag{13}$$

where $x = \{x_{inf}, x_{reg}, x_{cv}\}$ is the vector of model states ($x \in R^{20}$) and $\theta = \{\theta_{inf}, \theta_{reg}, \theta_{cv}\}$ is the vector of parameters ($\theta \in R^{88}$). The subscripts *inf*, *reg*, and *cv* represent the inflammatory, regulatory, and cardiovascular sub-models, respectively. That is,

$$x_{inf} = \{E, M_R, M_A, \text{TNF}, \text{IL6}, \text{IL8}, \text{IL10}\}$$
$$x_{reg} = \{T\text{emp}, \text{PT}, \text{NO}, R_s, H\} \tag{14}$$
$$x_{cv} = \{V_{la}, V_{lv}, V_{ao}, V_{vo}, p_{la}, p_{lv}, p_{sa}, p_{sv}\}.$$

The elements of $\theta$ are:

$$\begin{aligned}
\theta_{inf} = \{&k_E, k_M, k_{M\text{TNF}}, k_{MR}, k_{MA}, M_\infty, \eta_{ME}, h_{ME}, \eta_{M\text{TNF}}, h_{M\text{TNF}}, \eta_{M10}, h_{M10}, \\
&k_{\text{TNF}M}, k_{\text{TNF}}, \eta_{\text{TNF6}}, h_{\text{TNF6}}, \eta_{\text{TNF10}}, h_{\text{TNF10}}, w_{\text{TNF}}, \\
&k_{6M}, k_6, k_{6\text{TNF}}, \eta_{6\text{TNF}}, h_{6\text{TNF}}, \eta_{66}, h_{66}, \eta_{610}, h_{610}, w_{\text{IL6}}, \\
&k_{8M}, k_8, k_{8\text{TNF}}, \eta_{8\text{TNF}}, h_{8\text{TNF}}, \eta_{810}, h_{810}, w_{\text{IL8}}, \\
&k_{10M}, k_{10}, k_{106}, \eta_{106}, h_{106}, w_{\text{IL10}}\}
\end{aligned} \tag{15}$$

$$\begin{aligned}
\theta_{reg} = \{&\tau_1, k_T, k_{T\text{TNF}}, k_{T6}, k_{T10}, T_b, T_M, \eta_{T\text{TNF}}, \eta_{T6}, \eta_{T10}, h_{T\text{TNF}}, h_{T6}, h_{T10}, \\
&k_{\text{PTE}}, k_{\text{PT}}, PT_b,
\end{aligned}$$

$$k_{NM}, k_N, \eta_{NTNF}, h_{NTNF}, \eta_{N10}, h_{N10},$$
$$k_{RPT}, k_{RN}, k_R, \eta_{RPT}, R_b$$
$$\tau_2, k_H, H_b, H_M, BP_b, \eta_{HT}, h_{HT}, \eta_{HP}, h_{HP}\}$$
$$\theta_{cv} = \{R_{la}, R_{lv}, E_{la}, E_{lv}, E_{sa}, E_{sv}, V_{stroke}, E_m, E_M\}.$$

Table 1 lists all model parameters along with units and a description of their function.

**Nominal Parameter Values**

As described above, the model is comprised of three sub-models, namely the inflammatory, regulatory, and cardiovascular models. For each clinical study, we select a nominal parameter set which generates model dynamics that qualitatively align with the data trends.

*Inflammatory sub-model*: For Janum *et al*., parameter values for the inflammatory sub-model were taken from (7). For Copeland *et al*., parameters were adapted from (7), with several values being adjusted to achieve a better alignment between the model outputs and the data. Among the parameters that differ between the studies are the baseline levels of cytokines, the IL-6 synthesis in response to TNF-$\alpha$, the rate at which TNF-$\alpha$ activates monocytes, and the monocyte production rate of TNF-$\alpha$ and IL-6. See Table 2 for a complete list of the parameters different between the two studies.

*Cardiovascular sub-model*: For both studies we calculate the cardiovascular parameters as described in (6). The values for elastance in each compartment ($C_i$) and minimum and maximum elastance ($E_m$ and $E_M$), presented in Table 2, are computed with the following equations:

$$E_i = \frac{p_i}{V_{st,i}}, \qquad V_{st,i} = V_i - V_{un,i}, \qquad E_m = \frac{p_{lv}}{V_{ED}}, \qquad E_M = \frac{p_{la}}{V_{ES}}, \qquad (16)$$

where $V_{st,i}$ is the stressed volume and $V_{un}$ is the unstressed volume in compartment $i$. The end-diastolic and end-systolic volume in the heart are given by

$$V_{ED} = 142 - V_{un}, \qquad V_{ES} = 47 - V_{un},$$

with $V_{un} = 10$ (6).

The calculation of $V_{st,i}$ assumes that systemic volume is 85% of total blood volume, while the arterial and venous volumes are 20% and 80% of systemic volume, respectively (6). Total blood volume ($V_{tot}$) is a function of body surface area (BSA) and takes gender into account. They are given by

$$V_{\text{tot,female}} = 1000 \, (3.47 \text{BSA} - 1.954)$$

$$V_{\text{tot,male}} = 1000 \, (3.29\text{BSA} - 1.229)$$

$$BSA = \sqrt{\frac{h \, w}{3600}},$$

where $h$ (cm) is height and $w$ (kg) is weight (6).

*Regulatory sub-model*: For both studies, the parameters in the regulatory sub-system are adapted from (6) and adjusted where necessary to align with the data (Table 2). We set the parameters representing the baseline levels of temperature, HR, and BP ($T_b$, $H_b$, and $BP_b$, respectively) to the initial points in the data sets, that is, the measurements at $t = 0$. The equations for temperature and HR also have maximum level parameters ($T_M$ and $H_M$, respectively) which are informed by the literature. $T_M$ is 39.5° C, the cutoff for life-threatening hyperthermia (48), and $H_M$ is 207 (beats/min) (20). In addition, the delay $\kappa$ in the NO equation is chosen to ensure that an increase in NO is observed two to four hours after LPS administration (34).

**Parameter Estimation**

We fit the model to the measurements available for immune mediators, temperature, BP and HR, by minimizing the least squares error between the model output and the data averaged over the population of healthy subjects in each study. The least squares error $J$ is given by

$$J = r^T r, \qquad \text{where } r = \frac{1}{\sqrt{N}} \left( \frac{Y_{\text{model}} - Y_{\text{data}}}{\overline{Y_{\text{data}}}} \right), \tag{17}$$

where $Y_{\text{model}}$ and $Y_{\text{data}}$ are the model output and measured data, respectively. $\overline{Y_{\text{data}}}$ is the mean of the data, and $N$ is the total number of data points (38). For this study, we minimize $J$ using the built-in optimization routine *fminsearch* in MATLAB.

To determine which parameters to estimate, we used sensitivity analysis and subset selection for each sub-system of our mathematical model. Sensitivity analysis was carried out using a local derivative-based approach where parameters are varied one at a time to determine the sensitivity of the model output to its parameters. If small perturbations in a parameter result in significant changes in the output, the parameter is sensitive. If not, then the parameter is insensitive. The relative sensitivity was computed with a forward difference approximation, as described in (33) and ranked sensitivities were computed using the two-norm averaging time-varying sensitivities to find a group of sensitive parameters for each sub-model.

After a set of sensitive parameters was identified, we used subset selection by the correlation method outlined in (41) to determine which of the sensitive parameters were identifiable. A parameter is unidentifiable if it has a linear dependence on the values of other parameters. The parameters we estimated (given in Table 3) were ones that were both sensitive and identifiable. A detailed presentation of the sensitivity analysis and subset selection is given in (6) gives a detailed presentation of the sensitivity analysis and subset selection. Since the sensitivity matrix is evaluated at the nominal parameter values, this analysis is only valid in a neighborhood around the nominal values. Therefore, we repeated the subset selection at the optimized parameters to ensure that the estimated parameters are still identifiable.

**Therapeutic Interventions**

We use the physiological model to carry out an in-silico exploration of several therapeutic strategies for mitigating the harmful effects of an infection. Symptoms of an infection include elevated temperature, and in later stages, decreased blood pressure (22). Antibiotics are used to clear the pathogen from the body. Antipyretics help to decrease body temperature, and vasopressors increase organ bed resistance, leading to an increase in BP. In this light, our *in-silico* experiments focus on interventions with antibiotics, antipyretics, vasopressors, and a combination treatment which incorporates all three.

To model an infection state, we let the endotoxin to remain constant over a 12-hour time window. This approach simulates the situation when the body is not capable of clearing the pathogen from the bloodstream on its own (50). This causes the systolic BP to fall to a hypotensive level (below 90 mmHg) (42, 60). This state only arises after a prolonged period of inflammation not examined in the Copeland and Janum studies, who only considered the response to a bolus LPS injection resulting in controlled inflammation where systolic BP rises slightly in response to endotoxin and then decreases back toward the nominal level. By adding the simulation of infection, we show that our model can account for the physiological response during controlled inflammation as well as for the case of hypotension.

We captured the action of antibiotics by increasing the decay rate of LPS ($k_E$). The time frame for initiating antibiotic administration considers that medical intervention is not used before the infection has resulted in a significant rise in temperature. A commonly used indicator is fever, and the peak in body temperature occurs at about 4 hours. We modeled antipyretics by changing the PT and temperature dynamics, allowing them to reach baseline levels faster, and vasopressors by increasing the rate at which resistance approaches its baseline level, which in turn affects BP. Lastly, we look at a combined therapeutic protocol, that targets LPS, pain and fever, and resistance simultaneously. The Appendix gives a detailed description of the implementation of these therapeutic interventions.

**Results**

**Parameter Estimation**

We apply the physiological mathematical model to the two datasets. Fig. 5A shows model predictions together with the observed average behavior of immune mediators among human subjects exposed to endotoxin in the Copeland *et al*. study. Fig. 5C presents the comparison with the Janum *et al*. investigation. The pro-inflammatory cytokines TNF-$\alpha$, IL-6, and IL-8 start to increase about 1 hour following the LPS injection, reach a peak concentration at about 2 hours and return toward their baseline levels at about 4 hours. IL-10, the mediator with immunosuppressive activity, begins to rise shortly after the other cytokines, close to 1.5 hours post endotoxin administration. IL-10 reaches its maximum between 2.5 and 3 hours and causes the pro-inflammatory cytokines to decrease.

Figs. 5B and 5D present the comparison between the measured and model-predicted response in body temperature, HR, and BP following the LPS challenge. Fig. 5B is the model fit to the Copeland *et al*. data, and Fig. 5D to the Janum *et al*. data. The inflammatory cascade caused by endotoxin leads to a fever. The elevated TNF-$\alpha$ and IL-6 concentrations result in core body temperature significantly higher than the baseline between 3.5 and 4 hours. Subsequently, temperature causes a notable increase in HR within the same time frame. Also, the fever influences the interaction between BP and HR. LPS exposure leads to an increase in BP at approximately 1.5 hours, which acts to decrease HR, but the temperature effect dominates the dynamics, and HR remains elevated above baseline.

After fitting the model to data, we discovered that temperature alone is not enough to explain the HR dynamics. We observed that elevated temperature increases HR, but the subsiding of fever did not lead to a decrease in HR as seen in the data (Fig. 6). This indicates that, apart from temperature, there must be another HR control mechanism causing the restoration of homeostasis, and our findings suggest that BP could be this control mechanism.

We first tested an equation structure for the HR variable, where temperature and BP were accounted for as independent effects, by adding their contributions. The results (Fig. 6) reveal that this model does not fully capture the dynamics of the observed HR response. This led us to hypothesize that the effects of BP on HR, on the time scale of hours, are temperature-dependent, and the model fits to the data confirm this hypothesis.

## FIG.  5 – MODEL FITS TO DATA

## FIG.  6 – TEMP ALONE & INDEPENDENT TEMP AND BP EFFECTS

**Therapeutic Interventions**

We further probe the validity of the mathematical model by using it to predict responses to treatments. We examine the model predictions in the case of a sustained inflammatory event causing an infection as well as the effects of different treatment alternatives, including antibiotics, antipyretics, vasopressors, and a therapy combining all three medications. Results of these simulations are summarized in Figs. 7 and 8.

*Infection State*

Infection can be described by a sustained inflammatory event where the immune cells are ineffective at clearing it from the bloodstream. The physiological response to an untreated infection includes a significant increase in HR and a drop in BP, which in severe cases fall to a hypotensive level (defined as a systolic BP at rest below 100 mmHg).

An infection dramatically decreases the number of resting monocytes, leading to a higher level of activated monocytes, compared to the case of a controlled inflammatory response (Fig. S1). The pro-inflammatory cytokines peak earlier in the presence of an infection when compared to baseline. However, they reach similar maximum concentrations. The anti-inflammatory cytokine IL-10 reaches a notably higher level, and this translates into a slightly lower peak in temperature since IL-10 suppresses the thermal response. The temperature does not return to its nominal level as it does in a normal inflammatory response (Fig. 7) due to the slow decay of pyrogens.

The sustained presence of LPS during infection also causes a significant drop in the pain threshold. NO concentrations are higher due to release from the increased number of activated monocytes (Fig. 7). These response indicators have opposite effects on resistance. The initial change in PT is more dramatic than the increase in NO, resulting in a more significant increase in resistance when compared to the response in the absence of infection. However, after 4 hours, the rate of PT decline is slow, while NO is still sufficiently elevated. This results in a decreasing resistance. The rise in resistance leads to a higher BP than in the case of a normal inflammatory response, including the maximum value. As the resistance drops, a notable decrease occurs in BP, bringing it in a hypotensive range, which persists.

In response to an infection, our simulation results (Fig. 7) show a significant increase in HR as a temperature rises. The maximum HR reached is like that in a normal inflammatory response. As the thermal response subsides, HR begins to decrease with further downregulation from BP. However, as BP decreases into the hypotensive range, HR increases again to compensate for reduced blood circulation (23).

*Antibiotics*

In this scenario, the bacterial toxin has a sustained presence for 4 hours, and then antibiotics are administered to accelerate its clearance from the bloodstream. Antibiotics are a standard treatment used to fight bacterial pathogens (22, 50). Intervening with antibiotics prompts a slow recovery in the PT and helps to bring temperature and HR down toward their nominal levels. On the other hand, BP increases slightly, but there is no significant improvement toward recovery from the hypotensive range (Fig. 7). This is due to the introduction of antibiotics after the number of activated monocytes reaches a peak. The large cell population results in elevated NO, which has a stronger influence on the cardiovascular system compared to the change in PT.

*Antipyretics*

Fever is usually a part of the body's response to infection as well as an element in the inflammation from other causes. In the intensive care unit, about 70% of patients experience fever (11). In addition, surgical procedures are associated with a noticeable increase in body temperature (19). Antipyretic medication is a commonly used therapeutic approach for managing fever and pain to lessen the strain on the cardiovascular system and improve patient recuperation outcomes (53). As with antibiotics, introducing antipyretics into the model leads to recovery in HR. On the other hand, there is a moderate improvement in BP but only for a short time following antipyretic administration (Fig. 7). Then, BP falls to the hypotensive rage and does not recover. Also, the BP level decreases more compared to the case of antibiotic administration. This is due to NO being sufficiently elevated and exerting a stronger effect on peripheral vascular resistance than the change in PT. As antipyretics do not directly affect endotoxin levels, the inflammatory mediators and NO remain as in the infectious state. In addition, the decrease in HR causes further drop in BP due to lower cardiac output.

*Vasopressors*

The inflammatory response to an infection results in increased nitric oxide production (50), which leads to a significant drop in systemic blood pressures. Vasopressors may be used to elevate these pressures to a normal level (22, 49, 50). Like antipyretics, vasopressors do not affect endotoxin levels, resulting in no change in inflammatory mediators, NO, PT, or temperature. The action of vasopressors increases resistance, which translates into recuperation of systolic BP to its nominal level. Despite this, HR remains abnormally elevated (Fig. 7).

*Multimodal Treatment*

The multimodal therapeutic strategy simultaneously targeting toxin clearance, pain and fever, as well as low resistance, renders the best outcome since it brings about the alleviation of pain as well as recovery of normal body temperature, HR, and BP (Fig. 8).

**FIG. 7 – ANTIBIOTICS, ANTIPYRETICS, AND VASOPRESSORS**

**FIG. 8 – MULTIMODAL TREATMENT**

**Discussion**

The model successfully captures the immuno-pain-thermal-cardiovascular behavior and their interactions in two clinical studies of healthy adults under an LPS challenge. In response to an inflammatory reaction against a bacterial toxin, the model shows elevation of body temperature to a febrile level that we hypothesize leads to a notable rise in HR. These results are consistent with the data, Foteinou *et al*. (18) and our previous observations (7) that inflammation leads to an increased HR. While the study by Foteinou *et al*. hypothesized that the HR increase resulted from amplified efferent sympathetic activity and diminished parasympathetic activity, the study did not elucidate the role of temperature. Additionally, the study only indirectly accounted for the pain response through the inclusion of a sympathetic activity variable.

Inclusion of the thermal component into the current model illustrates the immune system's impact on HR via the febrile response. The release of inflammatory cytokines, in particular, IL-6 and TNF, raises core body temperature (61). Our findings suggest that the thermal response is an integral part of the interaction between BP and HR during an infection, on a time scale of hours, providing support for our hypothesis that during fever while BP is in a normotensive range, BP acts to lower HR, and when the temperature is at its baseline level, no dramatic HR fluctuations are observed.

Our model directly addresses the role of the pain pathway in the human response to endotoxin. Using data and observations from Janum *et al*. (31) our model reveals that the body's pain threshold diminishes during an endotoxin challenge. In addition, our investigation shows that the pain response is linked to cardiovascular dynamics and leads to a slight increase in BP in the early stages of the inflammatory reaction.

**Therapeutic Interventions**

The onset of an infection initiates a cascade of physiologic reactions acting at multiple timescales, complicating critical care for the patient. This study performs a simulation-based treatment investigation to predict the impact of several interventions commonly used in the clinical management of infection. We simulate an infection by keeping the endotoxin level constant, simulating the body's inability to effectively remove the endotoxin (50). The model predicts detrimental pathophysiological effects of sustained LPS presence including fever, increased sensitivity to pain, hypotension, and elevated HR. These results are consistent with the symptoms observed in sepsis (40). However, it should be noted that in septic patients who have a Gram-negative infection, the LPS level is much higher than the 2 ng/kg dose used in the clinical studies.

Antibiotics are typically used to treat bacterial infections. In the management of sepsis, broad-spectrum antibiotics are given to patients intravenously to combat the spreading pathogen and reduce the bacterial endotoxin level (22, 50). The simulation results from our model demonstrate that antibiotics lead to fever reduction, pain relief, and HR normalization, but not a notable improvement in BP.

Antipyretics are used to reduce core body temperature and pain (46). This treatment, however, does not block the dynamics of inflammatory cytokines (55). According to (46), the benefits and drawbacks of such medication are not well understood. Fever may serve an essential role in fighting the infection, but it may also negatively impact the cardiovascular system. Our findings indicate that while antipyretics favorably affect pain, body temperature, and HR, they are unable to reverse hypotension. Vasopressors are an alternative treatment option that targets the cardiovascular system. The simulation analysis illustrates that vasopressors are successful in normalizing systolic BP, but they fail to alleviate fever, pain, and reverse abnormally high HR.

Uncontrolled infection, which spreads and leads to sepsis, is a multifaceted condition with numerous possibilities for treatment, as seen in its pathophysiology and recent treatment guidelines (26, 27). Considering this, a multimodal treatment strategy was simulated, where antibiotics, antipyretics, and vasopressors are introduced simultaneously into the system. As the model captures complex interactions among immune, thermal, pain, and cardiovascular components, the goal was to assess how the combined administration of interventions impacts the predicted behavior and compares with the outcomes from the individual treatment administrations. Our results demonstrate that when combined, antibiotics, antipyretics, and vasopressors affect the model variables in essentially the same way as they did independently. Vasopressors act to restore normal BP, antipyretics restore PT and temperature, and the HR trends to its baseline value. Overall, this *in-silico* assessment shows that the combination treatment is the most effective way to relieve pain and bring about the recovery of normal temperature, HR, and BP. While these results are promising, future work is needed to understand how the medications act either synergistically or in competition to counteract an inflammatory event.

The data fitting and simulation analysis demonstrates that the model provides a platform to explore how specific biological pathways affect predicted cardiovascular outputs. There is an advantage in tracking the cardiovascular response through HR and BP since these quantities can be measured non-invasively and interpreted within minutes, allowing for a much quicker patient assessment compared to blood sampling to determine plasma levels of pro-inflammatory cytokines.

**Limitations and future work**

A limitation of our model is that it is calibrated over a short time course dictated by the time scale of the two endotoxin clinical studies, of duration 9 hours and 6 hours. This makes it difficult to assess its ability to track dynamics over days, which is needed to understand the response to an infection. Moreover, we only analyze data from a small number of subjects (10 and 20, respectively) increasing the uncertainty of model predictions.

In addition, there are several simplifying assumptions within the model. First, we assume that the endotoxin decays at a linear rate and is otherwise uncoupled from the rest of the model. This model is appropriate for studying the endotoxin response to a bolus LPS injection, but the model needs additional components to predict the response to treatment. In future studies, we will investigate the interactions among activated macrophages, pathogens and fever since there is evidence suggesting that aggressive treatment for body temperature higher than 38.5°C can negatively affect the recovery of critically ill patients, decreasing their ability to successfully clear infections (53). Furthermore, fever is beneficial in combating pathogens by limiting their reproduction, increasing the activity of many classes of antibiotics, and increasing the innate immune response (61).

Other potential extensions of the model include the investigation of treatments for sepsis. Our preliminary numerical experiments indicate that, qualitatively, antibiotic treatment slightly increases BP and lowers HR. Septic shock, a life-threatening condition, results from uncontrollable sepsis, and symptoms include low BP and rapid HR (40). Though our model is able to obtain accurate fits to the data and predict how an individual will respond to treatment, additional work is necessary to incorporate the effect of tissue damage on the inflammatory response, as shown in the model by Chow *et al*. (10).

Clinically, there are several evolving phases of infection from early to later stages, including sepsis, severe sepsis, and septic shock (22). These phases differ in several ways, including evidence of organ dysfunction and persistent hypotension (22). Our model currently does not distinguish between these phases, as it was developed to investigate dynamics within short time scales following an endotoxin challenge. To extend the model to analyze its response to an infection, the model should be solved over longer timescales (days).

**Conclusion**

This work develops the first physiological mathematical model that explains the interactions in humans among inflammation, body temperature, pain, HR, and BP. The model successfully captures the time-course of events and interplays observed during a pathogen challenge in healthy individuals reported in two clinical studies. For both studies, the inflammatory response is adequately controlled and contained, and the bacterial endotoxin is effectively cleared from the bloodstream within 6 to 9 hours. Simulation analysis of therapeutic interventions suggests that an untreated, sustained pathogen challenge would bring about abnormally elevated HR and low BP. Simulation results show that to remedy these detrimental effects, the most effective approach is to administer a multimodal treatment combining antibiotics with antipyretics and vasopressors to simultaneously target toxin, pain, fever, and low vascular resistance.

**Acknowledgements**

We would like to thank S. Janum, K. Møller, and S. Brix for their contributions to the data collection process.

**Grants**

Dobreva and Olufsen were supported in part by NSF–RTG–1246991. Puelz was supported in part by the NSF/DMS–RTG–1646339, and Brady was supported by The Jayne Koskinas Ted Giovanis Foundation for Health and Policy. Olufsen, Puelz, Larripa, and Brady were also supported by the Mathematics Research Communities via an award to the AMS (NSF–DMS–1321794).

## Appendix

## Inflammatory Response

The equations of the inflammatory response are given by

$$\frac{dE}{dt} = -k_E E.$$

$$\frac{dM_R}{dt} = -H_M^U(E)(k_M + k_{MTNF}H_M^U(TNF))H_M^D(IL10)M_R + k_{MR}M_R\left(1 - \frac{M_R}{M_\infty}\right)$$

$$\frac{dM_A}{dt} = H_M^U(E)(k_M + k_{MTNF}H_M^U(TNF))H_M^D(IL10)M_R - k_{MA}M_A$$

$$\frac{dTNF}{dt} = k_{TNFM}H_{TNF}^D(IL6)H_{TNF}^D(IL10)M_A - k_{TNF}(TNF - w_{TNF})$$

$$\frac{dIL6}{dt} = (k_{6M} + k_{6TNF}H_{IL6}^U(TNF))H_{IL6}^D(IL6)H_{IL6}^D(IL10)\,M_A - k_6(IL6 - w_{IL6})$$

$$\frac{dIL8}{dt} = (k_{8M} + k_{8TNF}H_{IL8}^U(TNF))H_{IL8}^D(IL10)\,M_A - k_8(IL8 - w_{IL8})$$

$$\frac{dIL10}{dt} = (k_{10M} + k_{106}H_{IL10}^U(IL6))\,M_A - k_{10}(IL10 - w_{IL10}),$$

(1)

where the up- and down-regulation of Y by X is modeled using the Hill functions, $H_Y^U(X) = \frac{X^h}{\eta_{YX}^h + X^h}$ and $H_Y^D(X) = \frac{\eta_{YX}^h}{\eta_{YX}^h + X^h}$, respectively. Full model details can be found in (7).

## Cardiovascular Response

By conservation of volume, the changes in volume in the large and small arteries and veins are

$$\frac{dV_{la}}{dt} = Q - q_a, \qquad \frac{dV_{sa}}{dt} = q_a - q_s, \qquad \frac{dV_{sv}}{dt} = q_s - q_v, \qquad \frac{dV_{lv}}{dt} = q_v - Q. \qquad (2)$$

The flow through each compartment is found using Ohm's law giving

$$q_{la} = \frac{p_{la} - p_{sa}}{R_a}, \qquad q_s = \frac{p_{sa} - p_{sv}}{R_s}, \qquad q_{lv} = \frac{p_{sv} - p_{lv}}{R_v}, \qquad (3)$$

where $R_a$ and $R_v$ are arterial and venous resistances, and $R_s$ is the peripheral vascular resistance.

For each cardiovascular compartment, the pressure and volume are related by

$$p_i - p_{tis} = E_i(V_i - V_{un}), \qquad (4)$$

where $V_i$ is the volume of the compartment, $V_{un}$ is the unstressed volume, $E_i$ is the elastance of the compartment, $p_i$ is the pressure in the compartment, and $p_{tis}$ is the tissue pressure.

**Therapeutic interventions**

We perform a theoretical therapy study over a 12-hour time window.

*Infection State*: To simulate an infection, we assume the amount of endotoxin ($E$) remains constant for 12 hours, i.e.

$$\frac{dE}{dt} = 0. \tag{5}$$

*Antibiotics* are administered four hours after the onset of the infection to mimic typical action of patients seeking help once they experience fever. We hypothesize that antibiotics will increase the rate at which the endotoxin clears from the bloodstream and model this as

$$\frac{dE}{dt} = \begin{cases} 0, & \text{if } t \leq 4 \\ -2 \cdot k_E E, & \text{if } t > 4. \end{cases} \tag{6}$$

That is, there is a constant, sustained endotoxin presence initially, which begins to decay once the antibiotics are introduced at $t = 4$. The rate at which this occurs is set to be twice as fast as the endotoxin would decay during an inflammatory event without intervention with an antibiotic.

*Antipyretics* are also introduced into the model at $t = 4$. We model the effect of antipyretics by changing the pain threshold (PT) and the temperature gain (Temp) making them approach their baseline values ($PT_b$ and $T_b$, respectively) at rates which are slightly faster than in the baseline simulation. To model this effect, we modify equations for pain perception and temperature as

$$\frac{d\text{PT}}{dt} = \begin{cases} -k_{\text{PTE}} E \, \text{PT} + k_{\text{PT}}(PT_b - \text{PT}), & \text{if } t \leq 4 \\ 2 \cdot k_{PT}(PT_b - \text{PT}), & \text{if } t > 4 \end{cases}$$

$$\frac{d\text{Temp}}{dt} = \begin{cases} \frac{1}{\tau_1}(-\text{Temp} + T_b + k_T(T_M - T_b)(k_{T\text{TNF}} H_T^U(|\text{TNF} - w_{\text{TNF}}|) \\ +k_{T6} H_T^U(|\text{IL6} - w_{\text{IL6}}|) - k_{T10}(1 - H_T^D(|\text{IL10} - w_{\text{IL10}}|))), & \text{if } t \leq 4 \\ \frac{2}{\tau_1}(-\text{Temp} + T_b), & \text{if } t > 4. \end{cases} \tag{7}$$

*Vasopressors* are also introduced 4 hours after the infection. Their action leads to an increase in peripheral vascular resistance ($R_s$). We model this effect by letting the peripheral vascular resistance approach its baseline value ($R_b$) more quickly, i.e.

$$\frac{dR_s}{dt} = \begin{cases} k_{RPT}\dfrac{\Gamma^2}{\Gamma^2 + {\eta^2}_{RPT}} - k_{RN}N - k_R(R_s - R_b), & \text{if } t \leq 4 \\ -2 \cdot k_R(R_s - R_b), & \text{if } t > 4. \end{cases} \tag{8}$$

Inflammatory-thermal- pain-cardiovascular response to a pathogen challenge

*Multimodal Treatment* also initiated 4 hours after the infection onset combines antibiotics, antipyretics, and vasopressors combining the models in (5-8).

**Tables**

**Table 1. Descriptions and units of model parameters (exponent (exp); in response to (irt); number of cells (noc); temperature (Temp); heart rate (HR); blood pressure (BP); nitric oxide (NO); organ bed resistance ($R_s$); pain threshold (PT)).**

| Par. | Description | Unit | Par. | Description | Unit |
|---|---|---|---|---|---|
| **Inflammatory Sub-Model** | | | | | |
| $k_E$ | Endotoxin decay rate | hr$^{-1}$ | $k_{MR}$ | $M_R$ proliferation rate | hr$^{-1}$ |
| $k_M$ | Rate that endotoxin activates monocytes | hr$^{-1}$ | $k_{MA}$ | Decay rate of $M_A$, TNF-$\alpha$, IL6, IL8, and IL10, respectively | hr$^{-1}$ |
| $k_{MTNF}$ | Rate that TNF-$\alpha$ activates monocytes | hr$^{-1}$ | $k_{TNF}$ | | |
| $k_{TNFM}$ $k_{6M}$ $k_{8M}$ $k_{10M}$ | $M_A$ production rate of TNF-$\alpha$, IL6, IL8, and IL10, respectively | $\dfrac{\text{pg}}{\text{mL hr noc}}$ | $k_6$ $k_8$ $k_{10}$ | | |
| $k_{6TNF}$ $k_{8TNF}$ | IL6 and IL8 synthesis in response to TNF-$\alpha$ | $\dfrac{\text{pg}}{\text{mL hr noc}}$ | $w_{TNF}$ $w_{IL6}$ $w_{IL8}$ $w_{IL10}$ | Baseline TNF-$\alpha$, IL6, IL8, and IL10, respectively | pg/mL |
| $k_{106}$ | IL10 synthesis in response to IL6 | $\dfrac{\text{pg}}{\text{mL hr noc}}$ | | | |
| $\eta_{ME}$ | $M_A$ half-max irt endotoxin | ng/kg | $h_{ME}$ | Exp. modulating $M_A$ effect on endotoxin | - |
| $\eta_{MTNF}$ $\eta_{6TNF}$ $\eta_{8TNF}$ | Half-max of $M_A$, IL6 and IL8, resp. regulating TNF-$\alpha$ | pg/mL | $h_{MTNF}$ $h_{6TNF}$ $h_{8TNF}$ | Exp. modulating $M_A$, IL6 and IL8, resp. effect on TNF-$\alpha$ | - |
| $\eta_{M10}$ $\eta_{TNF10}$ $\eta_{610}$ $\eta_{810}$ | Half-max of $M_A$, TNF-$\alpha$, IL6 and IL8, resp. regulating IL10 | pg/mL | $h_{M10}$ $h_{TNF10}$ $h_{610}$ $h_{810}$ | Exp. modulating $M_A$, TNF-$\alpha$, IL6 and IL8, resp. effect on IL10 | - |
| $\eta_{TNF6}$ $\eta_{66}$ $\eta_{106}$ | Half-max of TNF-$\alpha$, IL6 and IL10, resp. regulating IL6 | pg/mL | $h_{TNF6}$ $h_{66}$ $h_{106}$ | Exp. modulating TNF-$\alpha$, IL6 and IL10, resp. effect on IL6 | - |
| $M_\infty$ | Max number of monocytes | noc | | | |

**Table 1. (cont)**

| Par. | Description | Unit | Par. | Description | Unit |
|------|-------------|------|------|-------------|------|
| **Cardiovascular Sub-Model** ||||||
| $R_a$ $R_v$ | Resistance in arteries and veins, respectively | $\dfrac{\text{mmHg min}}{\text{mL}}$ | $E_{la}$ $E_{sa}$ $E_{lv}$ $E_{sv}$ | Elastance of large and small arteries and veins, respectively | $\dfrac{\text{mmHg}}{\text{mL}}$ |
| $E_m$ $E_M$ | Min and max elastance, respectively | $\dfrac{\text{mmHg}}{\text{mL}}$ |||||
| **Regulatory Sub-Model** ||||||
| $k_{TTNF}$ $k_{T6}$ $k_{T10}$ | Rate of Temp change irt TNF-$\alpha$, IL6 and IL10 | - | $\tau_1$ | Temp time constant | hr$^{-1}$ |
|||| $k_T$ | Temp rate of change | - |
| $\eta_{TTNF}$ $\eta_{T6}$ $\eta_{T10}$ | Half-max of TNF-$\alpha$, IL6 and IL10, resp. regulating Temp | pg/mL | $T_b$ $T_M$ | Baseline and max Temp, respectively | °C |
| $h_{TTNF}$ $h_{T6}$ $h_{T10}$ | Exp. modulating TNF-$\alpha$, IL6 and IL10, resp. effect on Temp | - | $k_{PT}$ | PT rate of change | hr$^{-1}$ |
|||| $k_{PTE}$ | Rate of PT change irt endotoxin | $\dfrac{\text{kg}}{\text{hr ng}}$ |
|||| $PT_b$ | Baseline PT | kPa |
| $k_{NM}$ | $M_A$ production rate of NO | (hr noc)$^{-1}$ | $k_N$ | NO decay rate | hr$^{-1}$ |
| $\eta_{NTNF}$ $\eta_{N10}$ | Half-max of TNF-$\alpha$ and IL10, resp. regulating NO | pg/mL | $h_{NTNF}$ $h_{N10}$ | Exp. modulating TNF-$\alpha$ and IL10, resp. effect on NO | - |
| $k_{RPT}$ $k_{RN}$ | Rate of R$_s$ change irt PT and NO | $\dfrac{\text{mmHg min}}{\text{mL hr}}$ | $\eta_{RPT}$ | Half-max of PT regulating R$_s$ | kPa |
|||| $R_b$ | Baseline R$_s$ | $\dfrac{\text{mmHg min}}{\text{mL}}$ |
| $k_R$ | R$_s$ decay rate | hr$^{-1}$ |||||
| $\eta_{HT}$ | Half-max of Temp regulating HR | °C | $\tau_2$ | HR time constant | hr$^{-1}$ |
| $\eta_{HP}$ | Half-max of BP regulating HR | bpm | $k_H$ | HR rate of change | - |
| $h_{HT}$ $h_{HP}$ | Exp. modulating Temp and BP, resp. effect on HR | - | $H_b$ $H_M$ | Baseline and max HR, respectively | bpm |
| $BP_b$ | Baseline BP | mmHg |||||

**Table 2. Nominal parameter values. All parameters of the inflammatory sub-model were taken from (7). All other parameters (cardiovascular and regulatory sub-models) were taken from (6). Parameters derived from data are denoted by ‡. Parameter denoted with ⚡ was taken from (20).**

| Par. | Value Janum *et al.* (2016) | Copeland *et al.* (2005) | Par. | Value Janum *et al.* (2016) | Copeland *et al.* (2005) |
|---|---|---|---|---|---|
| **Inflammatory Sub-Model** | | | | | |
| $k_E$ | 1.01 | | $k_{MR}$ | 0.006 | |
| $k_M$ | 0.041 | | $k_{MA}$ | 2.51 | |
| $k_{MTNF}$ | 4.14E-06 | 7.99E-06‡ | $k_{TNF}$ | 1 | |
| $k_{TNFM}$ | 0.60 | 0.49‡ | $k_6$ | 0.66 | |
| $k_{6M}$ | 0.81 | 0.92‡ | $k_8$ | 0.66 | |
| $k_{8M}$ | 0.56 | 0.34‡ | $k_{10}$ | 0.80 | |
| $k_{10M}$ | 0.0191 | | $w_{TNF}$ | 1.12‡ | 0.81‡ |
| $k_{6TNF}$ | 0.81 | 0.92‡ | $w_{IL6}$ | 1.05‡ | 1.74‡ |
| $k_{8TNF}$ | 0.56 | 0.34‡ | $w_{IL8}$ | 3.01‡ | 1.22‡ |
| $k_{106}$ | 0.0191 | | $w_{IL10}$ | 0.23‡ | 0.31‡ |
| $\eta_{ME}$ | 3.30 | | $h_{ME}$ | 1 | |
| $\eta_{MTNF}$ | 100.00 | 81.20‡ | $h_{MTNF}$ | 3.16 | |
| $\eta_{6TNF}$ $\eta_{8TNF}$ | 185.00 | 150.22‡ | $h_{6TNF}$ $h_{8TNF}$ | 2 / 3 | |
| $\eta_{M10}$ | 4.35 | | $h_{M10}$ | 0.30 | |
| $\eta_{TNF10}$ | 17.39 | | $h_{TNF10}$ | 3 | |
| $\eta_{610}$ | 34.77 | | $h_{610}$ | 4 | |
| $\eta_{810}$ | 17.39 | | $h_{810}$ | 1.5 | |
| $\eta_{TNF6}$ $\eta_{66}$ $\eta_{106}$ | 560.00 | 636.72‡ | $h_{TNF6}$ $h_{66}$ $h_{106}$ | 2 / 1 / 3.68 | |
| $M_\infty$ | 3E+04 | | | | |
| **Cardiovascular Sub-Model** | | | | | |
| $R_a$ | 0.19 | | $E_{la}$ | 0.81 | |
| $R_v$ | 0.0027 | | $E_{sa}$ | 3.85 | |
| $E_m$ | 0.0265 | | $E_{lv}$ | 0.0217 | |
| $E_M$ | 3.20 | | $E_{sv}$ | 0.13 | |

**Table 2. (cont.)**

| Par. | Value Janum *et al.* (2016) | Value Copeland *et al.* (2005) | Par. | Value Janum *et al.* (2016) | Value Copeland *et al.* (2005) |
|---|---|---|---|---|---|
| **Regulatory Sub-Model** | | | | | |
| $k_{TTNF}$ | 1.5 | | $\tau_1$ | 1 | |
| $k_{T6}$ | 1.5 | | $k_T$ | 0.5 | |
| $k_{T10}$ | 0.0625 | | $T_b$ | 36.83[‡] | 36.58[‡] |
| $\eta_{TTNF}$ | 185.00 | | $T_M$ | 39.5[§] | |
| $\eta_{T6}$ | 560.00 | | | | |
| $\eta_{T10}$ | 34.77 | | | | |
| $h_{TTNF}$ | 0.75 | | $k_{PT}$ | 0.08 | |
| $h_{T6}$ | 0.75 | | $k_{PTE}$ | 0.12 | |
| $h_{T10}$ | 1 | | $PT_b$ | 781.15[‡] | |
| | | | | | |
| $k_{NM}$ | 0.0020 | 0.0018 | $k_N$ | 0.045 | 0.025 |
| $\eta_{NTNF}$ | 95 | 70 | $h_{NTNF}$ | 2 | |
| $\eta_{N10}$ | 4 | | $h_{N10}$ | 0.4 | |
| | | | | | |
| $k_{RPT}$ | 30 | 14 | $\eta_{RPT}$ | 230.00 | 100.00 |
| $k_{RN}$ | 1.40 | 1.65 | $R_b$ | 1.03 | 0.99 |
| $k_R$ | 5.4 | 6.0 | | | |
| | | | | | |
| $\eta_{HT}$ | 36.5 | 36.39 | $\tau_2$ | 0.9 | 0.35 |
| $\eta_{HP}$ | 143.00 | 107.34 | $k_H$ | 0.21 | 0.20 |
| $h_{HT}$ | 2 | | $H_b$ | 60.36 | 64.29 |
| $h_{HP}$ | 4 | | $H_M$ | 207 | |
| $BP_b$ | 118.40 | 121.97 | | | |

**Table 3. Optimal parameter values.**

| Par. | Optimal Value | | Par. | Optimal Value | |
|---|---|---|---|---|---|
| | Janum *et al.* (2016) | Copeland *et al.* (2005) | | Janum *et al.* (2016) | Copeland *et al.* (2005) |
| **Inflammatory Sub-Model** | | | | | |
| $k_{10}$ | 0.83 | 0.65 | $k_{8M}$ | 0.46 | 0.15 |
| $k_{10M}$ | 0.01 | 0.01 | $k_{TNF}$ | 1.00 | 1.01 |
| $k_6$ | 0.81 | 1.02 | $k_{TNFM}$ | 0.60 | 0.56 |
| $k_{6M}$ | 0.72 | 0.93 | $k_E$ | 1.01 | 1.02 |
| $k_8$ | 0.74 | 1.21 | | | |
| **Cardiovascular Sub-Model** | | | | | |
| $R_b$ | 1.03 | 0.99 | $E_{la}$ | 0.79 | 0.80 |
| **Regulatory Sub-Model** | | | | | |
| $\tau_1$ | 1.38 | 2.12 | $\tau_2$ | 0.79 | 0.32 |
| $k_{T6}$ | 2.19 | 2.41 | $k_H$ | 0.20 | 0.20 |
| $k_{RPT}$ | 12.97 | 13.18 | $k_{RN}$ | 0.80 | 1.61 |
| $k_R$ | 4.28 | 6.60 | $k_{PTE}$ | 0.12 | - |
| $k_{PT}$ | 0.06 | - | | | |

**Figure Captions**

Fig 1. Experimental Protocol. Immune mediators (TNF-$\alpha$, IL-6, IL-8), temperature, heart rate, and blood pressure were periodically collected during the (A) Copeland and (B) Janum studies. Note that pain perception threshold and IL-10 were only recorded in the study by Janum *et al*. and heart rate was continuously recorded.

Fig 2. Feedback diagram for human response to endotoxin challenge. LPS administration initiates an immune cascade, as well as a decrease in the pain perception threshold. The decrease in pain perception threshold results in an increase in blood pressure (BP). Pro-inflammatory cytokines act as pyrogens, increasing body temperature while anti-inflammatory cytokines act as anti-pyrogens to decrease temperature. Pro- and anti-inflammatory cytokines have opposing effects on nitric oxide production, which decreases BP via vasodilation (decreases vascular resistance). Temperature increases heart rate (HR). BP decreases HR, via changes in vagal tone, in the case of normotension. When BP falls to a hypotensive range, it will act to increase HR. The changes in HR via BP are temperature-dependent.

Fig 3. Immune interactions in response to endotoxin challenge. Endotoxin (E) administration results in the activation of monocytes (MR→MA). Activated monocytes (MA) secrete mediators that induce further immune activation (TNF-α, IL-6, and IL-8). These pro-inflammatory mediators stimulate the production of IL-10, which regulates the immune response as an anti-inflammatory mediator. IL-6 also exhibits anti-inflammatory effects as it downregulates the synthesis of TNF-α and its own release.

Fig 4. Cardiovascular model. The cardiovascular system is comprised of the small and large arteries and veins (subscripts sa, la, sv, lv). Each compartment has an associated blood pressure $p$ (mmHg), volume $V$ (mL), and elastance $E$ (mmHg/mL). Flow between compartments are represented by $q_i$ (s/mL), with a corresponding resistance $R_i$ (mmHg s/mL) with subscripts (a, s, v) representing arteries, veins, and peripheral vasculature.

Fig 5. Model fits to data. Fits to experimental data from study by Copeland *et al*. (A-B) and Janum *et al*. (C-D). Endotoxin (2 ng/kg) was administered at time $t = 0$ and inflammatory mediator

response, temperature, heart rate, and blood pressure were recorded over 9 hours (A-B) and 6 hours (C-D).

Fig 6. Effect on heart rate dynamics of temperature alone and of temperature and blood pressure as independent effects for the (A) Copeland and (B) Janum studies. Temperature alone (blue curves) results in an increase in heart rate, however it is not sufficient alone to return heart rate to its baseline level. Including temperature and blood pressure as independent effects (red curves) also does not fully capture the dynamics of the observed heart rate response.

Fig 7. Effect of antibiotics, antipyretics, and vasopressors on cardio-inflammatory response. After inducing an infection via a constant endotoxin (black curve) (A), interventions were simulated at $t = 4$ (red dashed line). The infection causes (B) a slight decrease in temperature and (C) a dramatic decrease in the pain perception threshold. (D) Nitric oxide rises in response to the infection and does not respond to any intervention. (E) The resistance increases in response to the infection, resulting in (F) an increase in blood pressure and a subsequent (G) decrease in heart rate. Antibiotics (blue curves) result in a decrease in the endotoxin, pain relief (increase in PT), fever reduction, and heart rate normalization (the increase in heart rate between $t = 4.5$ and $t = 5$ is due to blood pressure being in the hypotensive range during that time). Antipyretics do not affect the endotoxin; however, they induce pain relief and decrease fever and heart rate. They are unable to counter hypotension. Similar to antipyretics, vasopressors do not affect the endotoxin. While they do not alleviate fever, pain, or abnormally high heart rate, they are able to normalize blood pressure.

Fig 8. Effect of multimodal treatment on cardio-inflammatory response. (A) Administering multimodal treatment at $t = 4$ effectively relieves (B) pain and (C) allows temperature, (E) blood pressure, and (F) heart rate to return to their baseline levels.

Fig S1. Effect of infection on inflammatory system. Infection (black curves) results in an increase in (A) all inflammatory mediators and (B) a dramatic decrease in the resting monocytes, resulting in an increase in the active monocytes.

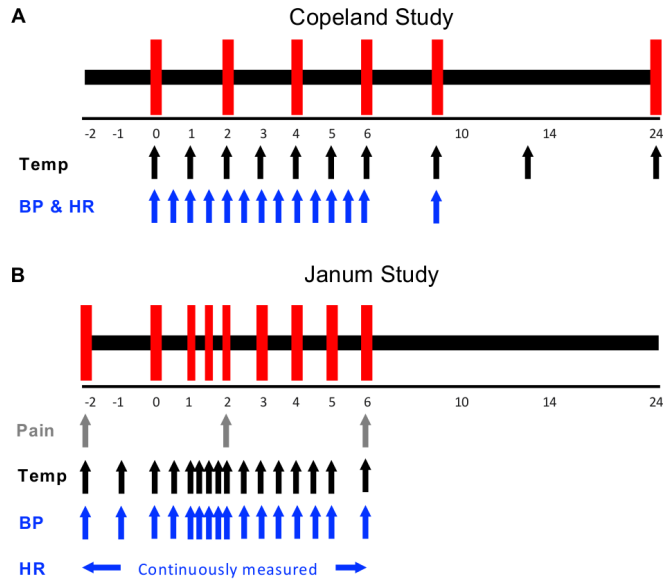

Fig. 1. Experimental Protocol. Immune mediators (TNF-$\alpha$, IL-6, IL-8), temperature, heart rate, and blood pressure were periodically collected during the (A) Copeland and (B) Janum studies. Note that pain perception threshold and IL-10 were only recorded in the study by Janum *et al.* and heart rate was continuously recorded.

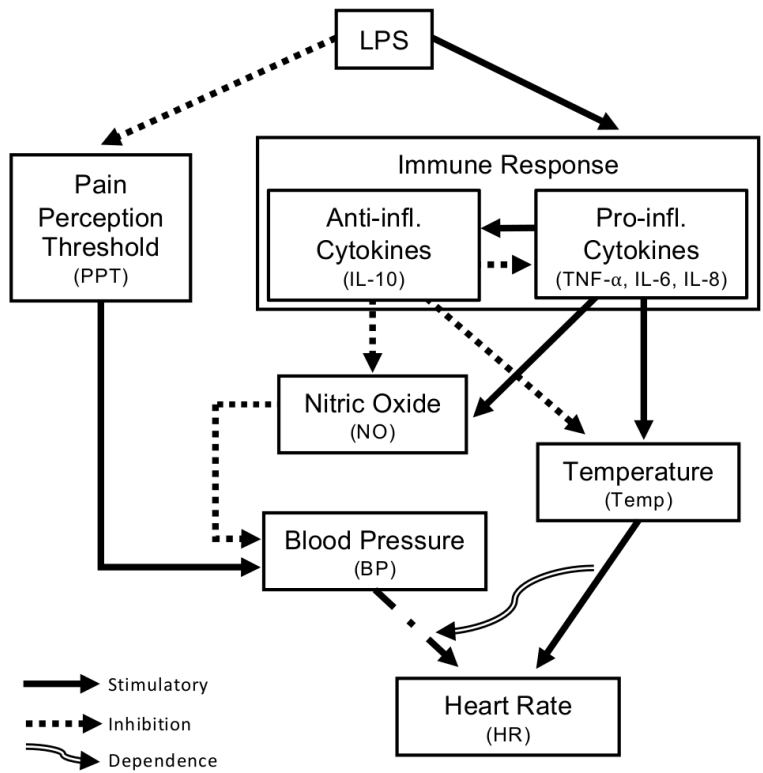

Fig. 2. Feedback diagram for human response to endotoxin challenge. LPS administration initiates an immune cascade, as well as a decrease in the pain perception threshold. The decrease in pain perception threshold results in an increase in blood pressure (BP). Pro-inflammatory cytokines act as pyrogens, increasing body temperature while anti-inflammatory cytokines act as anti-pyrogens to decrease temperature. Pro- and anti-inflammatory cytokines have opposing effects on nitric oxide production, which decreases BP via vasodilation (decreases vascular resistance). Temperature increases heart rate (HR). BP decreases HR, via changes in vagal tone, in the case of normotension. When BP falls to a hypotensive range, it will act to increase HR. The changes in HR via BP are temperature-dependent.

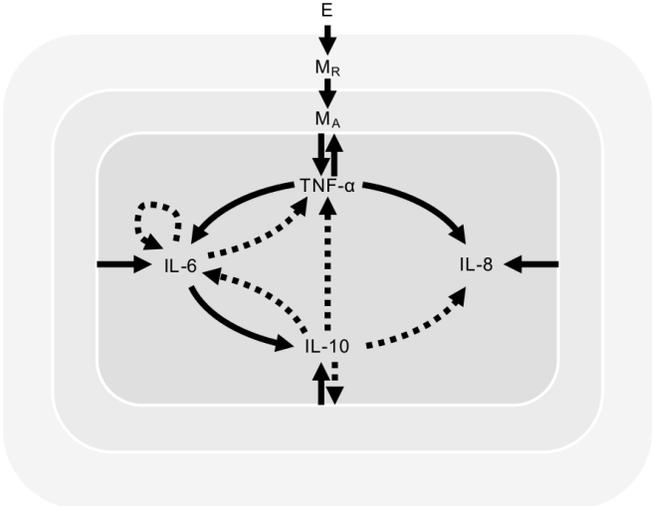

Fig. 3. Immune interactions in response to endotoxin challenge. Endotoxin (E) administration results in the activation of monocytes (MR→MA). Activated monocytes (MA) secrete mediators that induce further immune activation (TNF-α, IL-6, and IL-8). These pro-inflammatory mediators stimulate the production of IL-10, which regulates the immune response as an anti-inflammatory mediator. IL-6 also exhibits anti-inflammatory effects as it downregulates the synthesis of TNF-α and its own release.

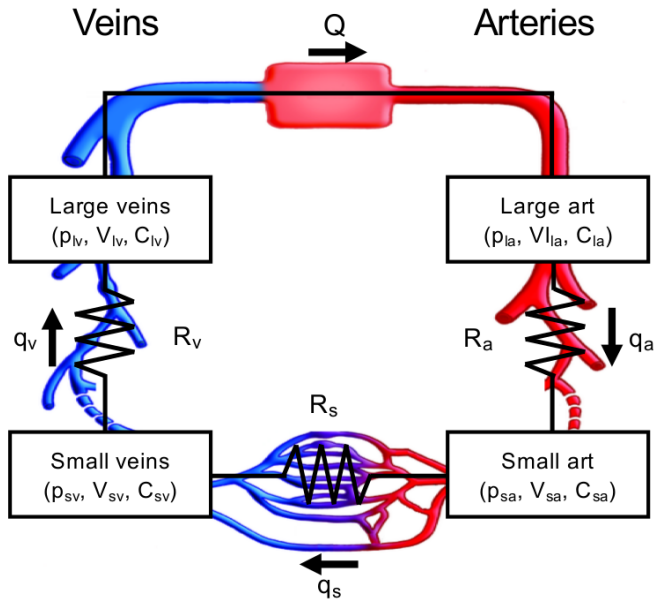

Fig. 4. Cardiovascular model. The cardiovascular system is comprised of the small and large arteries and veins (subscripts sa, la, sv, lv). Each compartment has an associated blood pressure $p$ (mmHg), volume $V$ (mL), and elastance $E$ (mmHg/mL). Flow between compartments are represented by $q_i$ (s/mL), with a corresponding resistance $R_i$ (mmHg s/mL) with subscripts (a, s, v) representing arteries, veins, and peripheral vasculature.

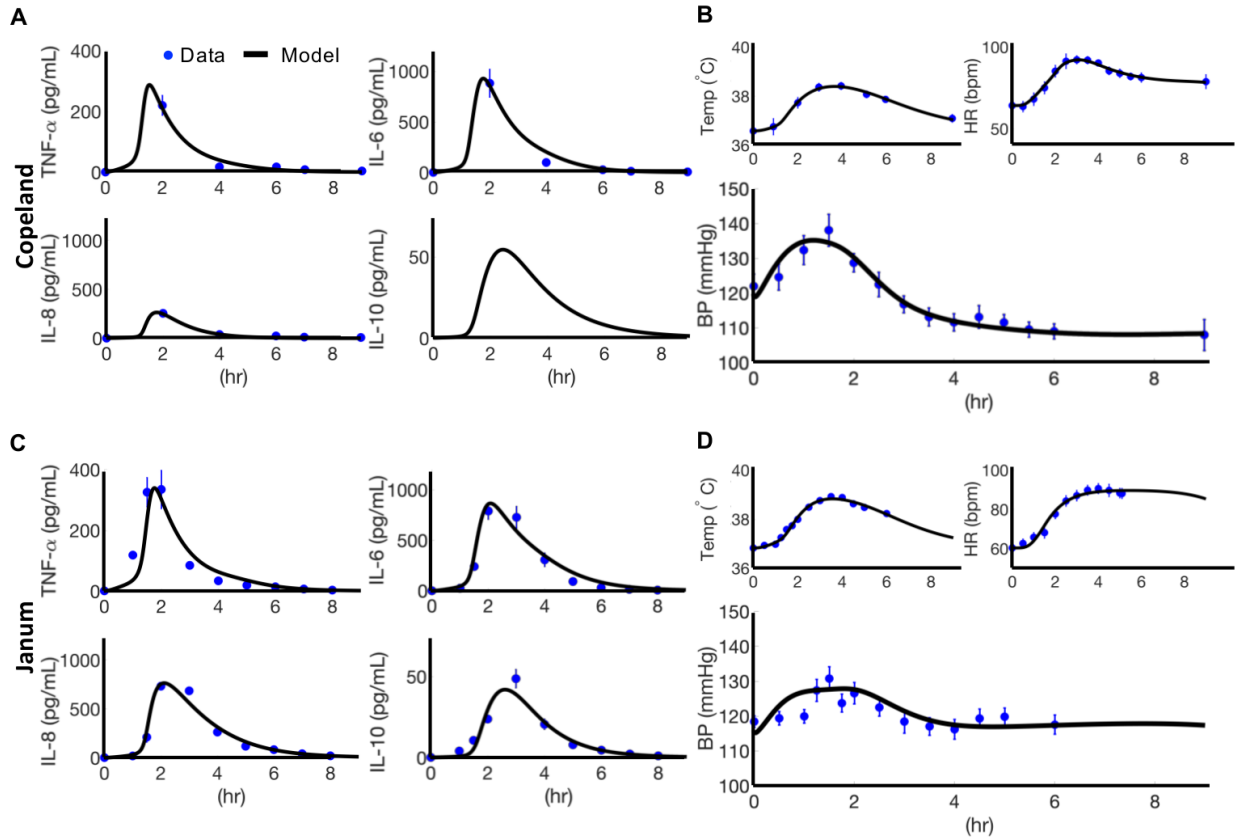

Fig. 5. Model fits to data. Fits to experimental data from study by Copeland *et al.* (A-B) and Janum *et al.* (C-D). Endotoxin (2 ng/kg) was administered at time $t = 0$ and inflammatory mediator response, temperature, heart rate, and blood pressure were recorded over 9 hours (A-B) and 6 hours (C-D).

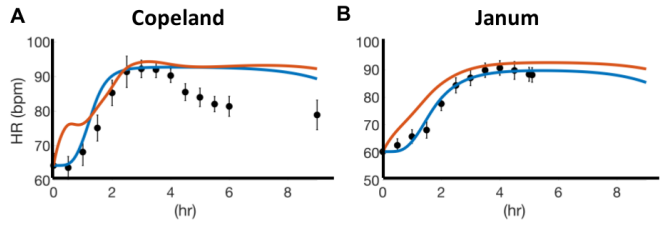

Fig. 6. Effect on heart rate dynamics of temperature alone and of temperature and blood pressure as independent effects for the (A) Copeland and (B) Janum studies. Temperature alone (blue curves) results in an increase in heart rate, however it is not sufficient alone to return heart rate to its baseline level. Including temperature and blood pressure as independent effects (red curves) also does not fully capture the dynamics of the observed heart rate response.

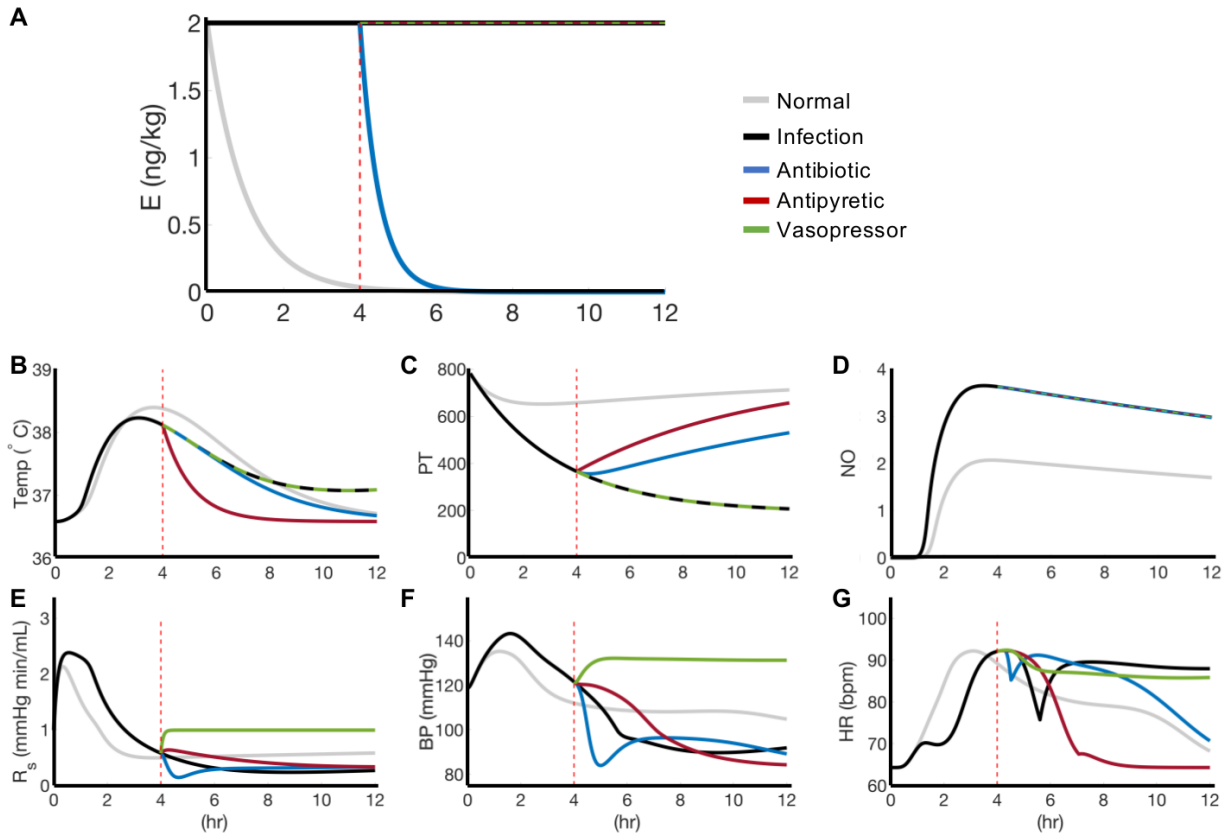

Fig. 7. Effect of antibiotics, antipyretics, and vasopressors on cardio-inflammatory response. After inducing an infection via a constant endotoxin (black curve) (A), interventions were simulated at $t = 4$ (red dashed line). The infection causes (B) a slight decrease in temperature and (C) a dramatic decrease in the pain perception threshold. (D) Nitric oxide rises in response to the infection and does not respond to any intervention. (E) The resistance increases in response to the infection, resulting in (F) an increase in blood pressure and a subsequent (G) decrease in heart rate. Antibiotics (blue curves) result in a decrease in the endotoxin, pain relief (increase in PT), fever reduction, and heart rate normalization (the increase in heart rate between $t = 4.5$ and $t = 5$ is due to blood pressure being in the hypotensive range during that time). Antipyretics do not affect the endotoxin; however, they induce pain relief and decrease fever and heart rate. They are unable to counter hypotension. Similar to antipyretics, vasopressors do not affect the endotoxin. While they do not alleviate fever, pain, or abnormally high heart rate, they are able to normalize blood pressure.

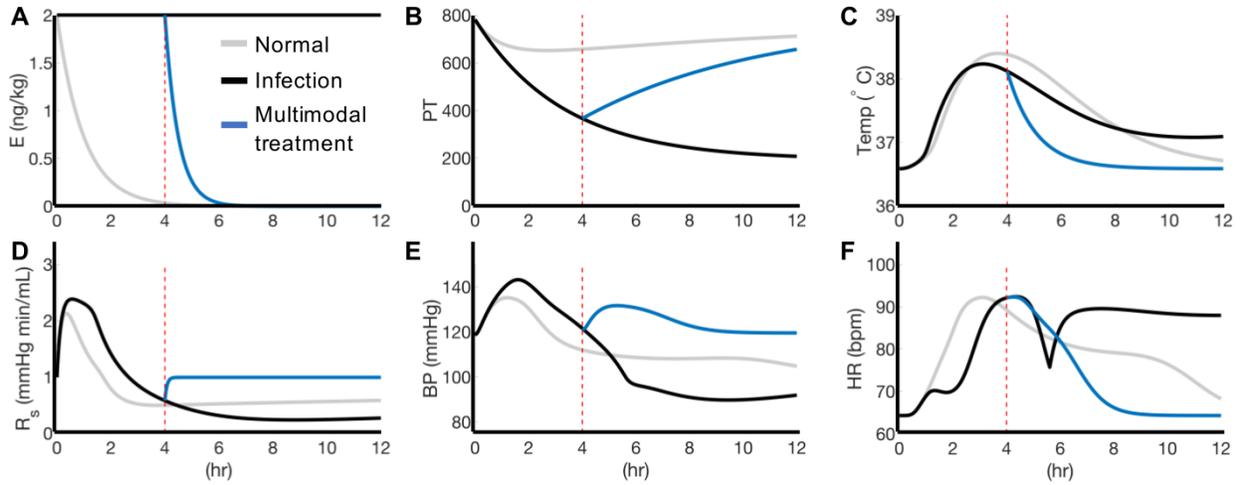

Fig 8. Effect of multimodal treatment on cardio-inflammatory response. (A) Administering multimodal treatment at $t = 4$ effectively relieves (B) pain and (C) allows temperature, (E) blood pressure, and (F) heart rate to return to their baseline levels.

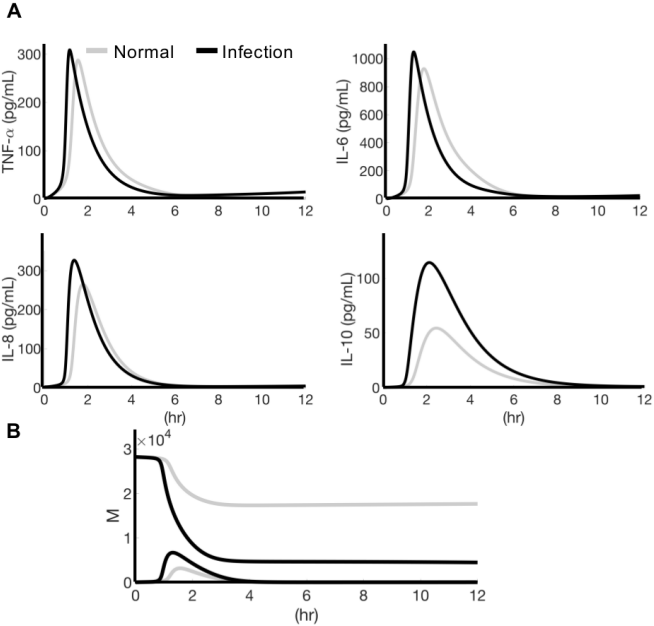

Fig. S1. Effect of infection on inflammatory system. Infection (black curves) results in an increase in (A) all inflammatory mediators and (B) a dramatic decrease in the resting monocytes, resulting in an increase in the active monocytes.